\documentstyle[11pt]{article}

\begin{document}
%

\begin{center}
{\large \bf Formation of partially dark-matter galaxies}

\vskip.5cm

W-Y. Pauchy Hwang\footnote{Correspondence Author;
 Email: wyhwang@phys.ntu.edu.tw} \\
{\em Asia Pacific Organization for Cosmology and Particle Astrophysics, \\
Institute of Astrophysics, Center for Theoretical Sciences,\\
and Department of Physics, National Taiwan University,
     Taipei 106, Taiwan}
\vskip.2cm


{\small(March 6, 2012; Revised: August 25, 2012; July 24, 2013; January 13, 2016)}
\end{center}

\begin{abstract}
The problem of galactic formation and evolution should be
solved on the basis of the Standard Model of particle physics.
We believe that we live in the quantum 4-dimensional
Minkowski space-time with the force-fields gauge-group structure
$SU_c(3) \times SU_L(2) \times U(1) \times SU_f(3)$ built-in
from the very beginning, i.e., the "background" of our world.
From this "background", we can see the lepton world, of atomic
sizes, and also the quark world, of $(fermi)^3$ sizes.
Basing on this belief, we study galactic formation and
evolution, concluding that our Cosmos should end up with
"model" galaxies. A model galaxy is the one in which a
spiral visible ordinary-matter galaxy, such as our
Milky Way and satellites, is surrounded by a huge
dark-matter neutrino halo. As a byproduct (of studying 
the Standard Model), we find that neutrinos 
will be the {\it only long-lived} dark-matter particles.

\bigskip

{\parindent=0pt PACS Indices: 05.65.+b (Self-organized systems); 98.80.Bp
(Origin and formation of the Universe); 12.60.-i (Models beyond the standard
model); 12.10.-g (Unified field theories and models).}
\end{abstract}

\section{Prelude}

{\it The galaxies in our Universe must be originated from the
Standard Model of particle physics. The galaxy, including its
invisible (dark-matter) part, should be derivable from the
Standard Model.}

Nowadays we have no difficulty to observe the various (visual)
ordinary-matter galaxies; but since the turn of the 21st century,
we realize that the Universe has 25\% in content in dark matter
while only 5\% in ordinary matter. The dark-matter world, which
in principle does not have strong and electromagnetic
interactions with us, is thus an integral part of our Universe.
Logic-wise, we so far always underestimate the importance of
those things which we could not see or could barely see.

On the other hand, our world should be described by the Standard
Model of particle physics, if the Standard Model is final, or
is almost final. For this, we advocate the Standard Model
\cite{Hwang417} that could explain the origin of mass
\cite{Origin}, that would explain the origin of fields
(point-like particles) \cite{Fields}, that could explain neutrino
oscillations in simple terms, that possesses $(123)$ in the quark
world while another $(123)$ in the lepton world. That is, our
world is the quantum 4-dimensional Minkowski space-time with the
force-fields gauge-group structure $SU_c(3) \times SU_L(2) \times
U(1) \times SU_f(3)$ built-in from the very beginning. From this
"background", we can see the lepton world, which is of the atomic
scale, and we also can see the quark world, typically of the
$(fermi)^3$ scale. Beyond the force-fields "background", the
lepton world, and the quark world, there is nothing more, i.e.,
we do not see anything else.

Thus, we should be able to describe our world, i.e.,
the surroundings, galaxies, etc., in a "complete" manner.
The description should be "complete", i.e., no left-over
as "unknowns". The various dark matters should be
included in the description - from the final Standard
Model of particle physics to the various galaxies. We
should not "blame" a large part of the Universe
{\it simply} to the $25\%$ dark matter.

{\it We know that the $3^\circ\,K$ cosmic microwave
background (CMB) exists in our Universe - an evidence
that the gauge-group structure is built-in from the very
beginning. We don't know if the predicted $1.9^\circ \,K$
cosmic background neutrinos/antineutrinos (CB$\nu$)
would be there, but the tiny mass of the neutrinos
should lead to the clustering phenomenon, that might
be observable eventually.}

\bigskip

\section{Super Phase Transition: The Origin of Mass}

All the masses come from the spontaneous symmetry breaking
(SSB) for the origin of mass \cite{Origin} - Before SSB, all
the mass terms are absent, except the "ignition" term, and,
after SSB, all the mass terms are generated through the
single symmetry breaking. Before SSB, the lepton world,
the quark world, and the "background" (i.e., the force
fields and the Higgs fields) are dimensionless, except
the "ignition" term. Because they are dimensionless, we
suspect that the values of all these couplings are
determined by the (quantum) 4-dimensional Minkowski
space-time, globally.

Maybe a suitable analogue of the mass SSB is the
phase transition of some sort. After the transition,
the "mass" assumes some meaning - some object has
the mass which can be measured. Before the SSB
transition, none of these guys (objects) has the
mass - or, the "mass" does not have the meaning.
We in fact struggle with the philosophical
meaning of the "mass".

There are two crucial implications for such mass
generation. First, before SSB (such as at a high
enough temperature, or, at the early Universe),
there are {\it no} mass terms - we describe a
world without mass terms. That is, we have to
learn to describe our world without the concept
of "mass". Second, apart from the "ignition"
term, all the couplings are dimensionless -
it is a little bit unthinkable conceptually
but it is true. Our world indeed was created by
the dimensionless God with a single "ignition"
key.

Since galactic formation is the mass accumulation
process, it does not have the meaning before SSB
has taken place. Thus, we label "the SSB resulting
in the origin of mass" as a super phase transition
-- the super phase transition after which all
particles gain masses (except the photons) and before
that the concept of "mass" simply has no meaning.
Thus, the beginning of galactic formation in our Cosmos
is when all the masses of the various particles
are generated via the SSB on the origin
of mass \cite{Origin}.

Our work on the origin of mass \cite{Origin} indicates
that the various masses would not be there if the temperature
is higher than the critical temperature $T_c$. Thus, the
evolution and formation of galaxies should be functions
of temperature, and it has meaning after it is lower
than the critical temperature $T_c$.

{\it Maybe it is important to name the phase transition
on the origin of mass \cite{Origin} as a "super
phase transition", though during these days we
use the concept of the spontaneous symmetry
breaking (SSB) to describe the phenomenon.
Clearly, the conceptual invention of the
spontaneous symmetry breaking, via complex
scalar Higgs field(s), is a magic.}

\bigskip

\section{What is the real content of the Galaxy?}

Since all the dark-matter particles in the Standard Model
\cite{Hwang417} will decay, sooner or later, into
neutrinos and antineutrinos \cite{Hwang2}, the story
on the formation and evolution of the various galaxies
could be rather simple.

Unlike the $3^\circ\,K$ Cosmic Microwave Background
(CMB) where photons are uniformly distributed in
space, the $1.9^\circ\,K$ Cosmic Background (CB)
neutrinos, owing to their masses, would {\it not}
distribute uniformly in space - leading to clusters
or halos. {\it Our Milky Way could offer us a
perfect model for the galactic formation - a visible
galaxy immersing in a large, but finite, halo of
neutrinos and antineutrinos.}

If we use the distance between the Milky Way and its
major satellite, the Large Magellan, $50\, kpc$, and
the visible Milky Way of $100,000$ light years, both
as the stick, we propose that the neutrino halo
encircles both of them, with a halo radius of, say,
$300,000$ light years, still far from the nearby
Andromeda Galaxy (at 2M light years). There are
voids in our Universe, each much bigger than the
Virgo Galaxy Cluster; judging from the biggest
neutrino mass of $0.058\, eV$, the existence of
much bigger voids may not be relevant for this
problem. So, the neutrino halo of radius
$300 kpc$, suitable for our Milky Way, would
be our example and it means that the core
neutrino density on the neutrinos cluster (halo)
of a visible ordinary-matter galaxy would be
enhanced by a factor of $10^5-10^{10}$, a really
conservative guess.

{\it For the sake of convenience, we will call
the galactic system of the above Milky Way and
its Satellites plus the much bigger neutrino
halo, with the radius about three times of the
visible part, as "the Model Galaxy".}

\bigskip

\section{Galactic Formation and Evolution}

{\it What would be the ingredients of aggregates of
stars and galaxies, as from the Standard Model
of particle physics \cite{Hwang417}? We should connect
the problem of galactic formation and evolution with
the knowledge of the Standard Model.}

If we advocate the "old" minimal Standard Model, then
we would believe that there would be something beyond
the Standard Model. The Standard Model which we would
advocate \cite{Hwang417} is a consistent, and likely
"complete", extension of the Model, that would explain
the origin of mass \cite{Origin}, that would
understand the origin of fields (point-like particles)
\cite{Fields}, that implements three generations of
point-like Dirac particles, and that explains
neutrino oscillations. In other words, we could use the
Standard Model of ours \cite{Hwang417} to enumerate
the ingredients, to start thinking about the
question of galactic formation and evolution.

In the origin of mass \cite{Origin}, the masses arise
from the spontaneous symmetry breaking (SSB), which we
also name it as a "super phase transition". The two
masses attract each other, so different from two
massless objects such as photons. It would be
interesting this aspect from general relativity;
before proving it, let's assume it as a basic
postulate.

If we could use Newton's gravitational law, the
attractive gravitational force exerted by the
core center, and thus the acceleration of the
object, is independent of the value of the object
mass, as long as the object has the mass. It does
not seem that, whether the object mass does
oscillate or not, the story would change much.
We may assume of the similar situation if we
switch from Newton's gravitational law to
Einstein's general relativity. Here we have to
note, from the independence of the object mass,
that the concept of "mass" is rather peculiar
since $m=0$ (no mass) differs from $m\to 0$
(vanishingly small mass).

As a parenthetical remark, we could advance
the concept of "mass" if, experimentally, we
could differentiate the clustering in the
case of CB$\nu$'s (having the tiny mass)
from the uniformity in the CMB (no mass).

Our attitude is as follows: The problem of galactic
formation and evolution, if addressed properly, should
be a highly academic, and challenging, research problem.
To begin with, we have to admit that we live in the
quantum 4-dimensional Minkowski space-time with some
force-fields gauge-group structure built-in from the
very beginning. The force-fields structure should be
built-in from outset since, otherwise, it is very
difficult to understand where it would come from.
The next stage of our thinking is that the Standard
Model \cite{Hwang417} is the consequence of the
space-time with some group structure built-in from
the outset. If we ignore the "ignition" term, the
whole theory, the Standard Model of \cite{Hwang417},
is dimensionless, i.e., that it does not have a
coupling with some dimension. Granting this Standard
Model, we can enumerate all the particles and their
lives - particles of the longer lifetimes would
aggregate, eventually forming stars, galaxies, etc.

So, the stable ingredients (for dead stars and
dead galaxies) include electrons
(positrons), protons, $\alpha$-particles ($^4He$
nuclei), $\alpha$-nuclei (such as $^{12}C$,
$^{28}Si$, $^{56}Fe$, etc.), photons such as
the $3^\circ K$ cosmic microwave background
(CMB), and neutrinos such as the $1.9^\circ K$
cosmic background (CB) neutrinos. Of course,
the fraction of those dead stars containing
$\alpha$-nuclei would be rather small.

This may sound very different from the
Standard Model \cite{Hwang417} but it is
not the case if we understand from the
publication of the Particle Data Group
\cite{PDG}. The point is that all heavy
particles decay in no time, i.e., shorter
that a $\mu Sec$. So, we have, as the
above, the list of the stable ingredients
for stars and galaxies.

In our Universe, we know that the CMB photons are
uniform - serving as an evidence that massless
particles do not cluster. On the other hand, at
least one species of neutrinos has a mass of
$0.058\,eV$, several orders bigger than
$1.9^\circ K$. Note that $1\,kT= 8.617\times 10^{-5}\,
eV$. Thus, the clustering of neutrinos is of
critical importance, leading to the model
galaxies as mentioned before.

{\it This is how we formulate the problem of
galactic formation and evolution in the Standard
Model \cite{Hwang417}. Or, in simple terms, it's
"the problem of the SM galactic formation and
evolution". It would be better to stick to
the same theoretical framework
if one is trying to criticize our solutions
of the problem.}

In the origin of mass \cite{Origin}, it is of
critical importance that the spontaneous symmetry
breaking (SSB) is responsible for the various
masses. Before SSB, there is no mass term anywhere
in the lagrangian - that is, all mass terms are
generated from SSB. Galactic formation is a
physical process of mass accumulation. So, there
is no galactic formation, before SSB had taken
place in the early Universe.

\medskip

\parindent=0pt

{\it What is the origin of the aggregation
into the clusters?}

\parindent=12pt

The phenomenon of clustering, e.g., in our case formation of
ordinary-matter galaxies and, possibly, of dark-matter
galaxies, is in fact a highly nonlinear effect. We do know
that the cluster would attract other to make it bigger,
i.e., to grow. In Newton's law, this is trivial and could be
understand as a "mass" effect. As argued in the above, CMB
is highly uniform while the CB neutrinos are believed to
be clustered because of the $0.058\,eV$ mass. But, in
Einstein's general relativity, we still don't know how
to perceive this aspect.

Like the grown-up process, clustering seems to have a life of certain
form on its own. This is particularly the case when we talk about
the seeded clustering, such as, the aggregation of the matter
at the early stage (producing the seeds via strong and
electromagnetic forces) followed by the clustering via gravitational
force. It is essential to recognize the processes of producing of
heavy $\alpha$-nuclei which then serve the so-called "seeded
clustering" - thus, the gravitational force is {\it not} the only
reason for clustering.

It is well-known that the rotation curve of our own spiral galaxy,
the Milky Way, is dragged by something invisible that may be four or
five times the mass of the Milky Way, if Newton's gravitational law or
Einstein's general relativity is assumed to be valid. This is a natural
candidate for the clustered dark matter. It is well-known that dark
matter occupies 25\% of the current Universe, as oppose to 5\% of
the visible ordinary matter. And we know that dark matter, owing to
their masses, should clusterize even due to gravitational forces,
except whether it was soon enough, as compared to one Giga years,
the time span that ordinary-matter galaxies form.

Galactic formation (for visible galaxies), since its start
from baryons, nuclei, atoms, molecules, complex molecules, and
chunks of matter, those from strong forces, electromagnetic
forces, and their residual forces, apparently is influenced
greatly from something that is not gravitational. We refer
this case as the "seeded" clustering. If the "seeded"
clustering of this kind were absent, whether galaxies
could exist in the time span of the age of the young
Universe, say, $10^9$ years, is, in our opinion, fairly
questionable.

As mentioned earlier, the spiral arm of our own galaxy, the
Milky Way, is influenced by the dark-matter clouds of (4-5) times of
the mass of the Milky Way, judging from the unusual tail (larger)
velocity, or the rotation curve, of the arm. The evidence seems to be
reasonably good also for other spiral galaxies but of course it would be
very interesting if it could be further substantiated. This is why
we try to study the so-called "model galaxy".

In the ordinary-matter world, strong and electromagnetic forces make the
clustering a very different story - they manufacture, via electromagnetic
or strong forces, atoms, molecules, complex molecules, and chunks of matter,
and then clustering gravitationally into the stars and the galaxies; the
so-called "seeded clusterings".

In the Standard Model \cite{Hwang417}, we realize that there are only
three complex scalar (Higgs) fields: $\Phi(1,2)$ (the SM Higgs),
$\Phi(3,2)$ (the mixed family Higgs), and $\Phi(3,1)$ (the purely
family Higgs), and they close the Higgs sector nicely, making family
gauge bosons and the weak bosons massive. The Higgs
particles, the family gauge bosons, and the weak bosons have
the masses in the range between a few $GeV$ and about $100\,
GeV$, with the decal channels containing neutrinos/antineutrinos.
It is not possible to produce the seeds from the dark-matter
world.

Perhaps we have to question whether the Standard Model of
\cite{Hwang417} is unique. We mention that, apart from the
"ignition" term, it is a dimensionless theory, which should
be determined by the quantum 4-dimensional Minkowski
space-time, {\it not} by the individual object inside.
Hence, the chance of the uniqueness of the Standard
Model in fact would be better than the old Standard Model.

What we have in mind is that, in the ordinary-matter world, the sequence
of atoms, molecules, complex molecules, etc. (up to the mass of one
$TeV$ level) yields the "seeds" of the clusterings - the very essential
"seeded" clustering of all time, the seeded clustering that is
relevant for the time span of our young Universe. Whether there are
such seeds for the dark-matter world is a very essential question.
For our world, i.e., the quantum 4-dimensional Minkowski space-time
with the force-fields gauge-group structure $SU_c(3) \times SU_L(2)
\times U(1) \times SU_f(3)$ built-in from the very beginning, the
Standard Model \cite{Hwang417} is in fact {\it the consequence}.
The family gauge bosons and the family Higgs particles are the
{\it only} unseen particles - they come from the gauge
group $SU_f(3)$ and, because of the massiveness, from the
complex scalar Higgs fields, $\Phi(3,2)$ (the mixed family Higgs)
and $\Phi(3,1)$ (the purely family Higgs). These particles
decay mostly to neutrinos and antineutrinos, with $sub-GeV$
widths (very similar to the cases of $W^\pm$ and $Z^0$).

Thus, {\it in our world, there are no new seeds coming from
the dark-matter sector for the seeded clustering.} This
conclusion comes from the simplicity of the Standard Model
\cite{Hwang417} for our world. Note that neutrinos (the known
one kind of dark matter) alone, with their tiny masses, can
in principle aggregate but over the time span much longer than the age of
our Universe, $13.7\, Gyr$. The ordinary-matter galaxies are manufactured
in about the first $1\, Gyr$, but clearly with the help of the "seeded"
clusterings, as mentioned above.

Theoretically we could start with the Standard Model of particle physics
\cite{definition}, which specifies, maybe initially, the ordinary-matter
world, and describe all kinds of the known interactions - strong,
electromagnetic, weak, and family interactions. These particles do
aggregate (into atoms, molecules, and then macroscopically
gravitational objects) under strong, electroweak, and
gravitational forces (in a time span of $1\, Gyr$, the life
of the young Universe). {\it The question of whether there are
essential dark-matter particles serving as the new seeds for
the clusterings, in turn for galactic formation and
evolution, is in fact tied to the validity of the
Standard Model. In our Standard Model \cite{Hwang417},
the answer is "no" - that is, there is no dark-matter
seed for clustering.}

\bigskip

\section{The Perspectives as from the Standard Model}

If the solution to the problem of galactic formation and evolution
could be closely linked to the Standard Model (of particle physics),
then both the problem (of galactic formation and evolution) and the
Standard Model would make a lot of sense. It is in this direction,
in which we think that we are making progresses.

As mentioned in our discussions of the origin of mass and subsequently
on the origin of fields (point-like particles) \cite{Origin, Fields},
we assert that our world is the (quantum) 4-dimensional Minkowski
space-time with the force-fields gauge-group structure $SU_c(3)
\times SU_L(2) \times U(1) \times SU_f(3)$ built-in from the very
beginning. We see the lepton world, of the atomic sizes, and we
also see the quark world, of the $(fermi)^3$ sizes. Both worlds
are of the $(123)$ gauge symmetry and thus they are well-behaved
in the limit $r\to 0^+$. This offers us the framework for
discussing and analyzing (visible or invisible) galactic
formation and evolution.

We begin with the $3^\circ K$ cosmic microwave background (CMB) and the
$1.9^\circ K$ cosmic background (CB) neutrinos - CMB is found to be
rather uniform and photons are massless, while at least one species
of neutrinos are of the mass $0.058\, eV$ and CB neutrinos are supposed
to be clustered. {\it This is the environment for galactic formation and
evolution in our world. The existence of the $3^\circ\,K$ CMB is the
indication of why the gauge-group structure was in the Minkowski
space-time in the very beginning.}

The key difference between CMB and CB$\nu$ is the mass effect - photons
are massless but certain neutrinos have the tiny mass. Since certain
neutrinos have the tiny mass, they have to cluster. If we could use
Newton's law for gravitational force and then we could differentiate
between CMB and CB$\nu$, it is already decreed that Einstein's
general relativity is superior to Newton's. In reality, differential
geometry might not bring in too much unexpected in ordinary cases.
So, photons are massless and lack of mass explains why CMB is
virtually uniform, and {\it vice versa}. But CB neutrinos
must have clustered - in the core of neutrino halo, the
neutrino density could be $10^5$ times than that in the massless
case. That is why we try to study "model galaxy" for the
problem of galactic formation and evolution.

In the Standard Model \cite{Hwang417}, all the gauge bosons and the
associated Higgs decay into the lighter particles, in the multi-GeV
ranges. On the Dirac fermions, the heavy quarks decay into light
one ($u$ and $d$) while the heavy leptons into light ones ($e^\mp$
and some species of $\nu$). Protons, neutrons, atoms, and molecules
are the forms for the left-overs. On the gauge and Higgs bosons,
the photons will be the {\it only} left-overs. Thus, the CMB and
the CB$\nu$ would be there. For galaxies, these would be the
forming primary materials. Assuming the Standard Model
\cite{Hwang417}, these would be the {\it only} forming
primary dark-matter materials.

{\it In fact, we might be able to "prove" that the Standard Model
\cite{Hwang417} is indeed the final one in our World. Why is it so?}

No.1. We all agree that the relativity principle of Einstein and
the quantum principle are the two corner stones of modern physics
of the 20th Century. That is, we declare that we live in the quantum
4-dimensional Minkowski space-time.

No.2. We see that the $3^\circ K$ cosmic microwave background (CMB)
is there and it is rather uniform. There is no clustering effect
due to mass, indicating that photons are massless.

Because of No.2., we see the necessity of adding the force-fields
gauge-group structure to the space-time. We do it at the very
beginning, since there is no {\it a priori} priority between
the Lorentz group and the gauge group. This coincides the fact
that we see CMB.

No.3. The complex scalar field cannot {\it exist alone}
since it is self-repulsive due to $\lambda (\phi^\dagger \phi)^2$
(with $\lambda={1\over 8}$). The existence of the complex scalar
Higgs fields $\Phi(3,2)$ (the mixed family Higgs) and $\Phi(3,1)$
(the purely family Higgs) guarantees the existence of the
Standard-Model Higgs.

No.4. In describing the motion, Einstein's basic relation,
$E^2={\vec p\,}^2 + m^2$, serves the fundamental role. So is
Dirac's linearization, $E={\vec \alpha}\cdot {\vec p} +
\beta m$. That is, the Dirac equation applies for electrons,
neutrinos, quarks, etc.

No.5. If we write any other fields, they do not have a way
to interact with the Standard Model particles, since the SM
particles are consistent and closed in the logical sense, and
they don't need anything in extra.

No.6. Maybe the repetition from the lepton world to the quark
world could be possible, since we need to understand more the
play of scales to say something definitive.

We can see the lepton world, because of the $SU_L(2)
U(1) \times SU_f(3)$ symmetry. We can see the quark
world, in view of the $SU_c(3) \times SU_L(2) \times
U(1)$ symmetry. Because of the $(123)$ symmetry, the
"background" can see these different worlds. Different
$SU(3)$ corresponds the different scale - the quark
world and the lepton world have rather different, and
far apart, scales. But they have smooth behaviors in
the different limits, such as $r\to 0$, because of
$(123)$ symmetry.

{\it But how do we know that we have to stop with
the lepton world and the quark world, no more,
such as the pre-quark world at a much smaller
scale?}

Anyway, now we have in mind the Standard Model \cite{Hwang417},
to use it to analyze the problem of galactic formation and
evolution. The entire world given by the Standard Model, i.e.,
the force-fields "background" plus the lepton world and the
quark world, is relatively simple.

\bigskip

\section{Dirac Similarity Principle and Minimum Higgs Hypothesis}

{\it Where we do live is some place that satisfies both the relativity
principle and the quantum principle - i.e., the quantum
4-dimensional Minkowski space-time. In this world, Einstein
basic relation, $E^2= {\vec p\,}^2 + m^2$, plays the basic
role of the motion while the Dirac's linearization, $E=
{\vec \alpha}\cdot {\vec p} + \beta m$, is required for a more
intimate description of the motion.}

Maybe it is worthwhile to backtrack in saying a few words about "Dirac
Similarity Principle" and "Minimum Higgs Hypothesis", the two
working rules that have helped us to think forward over the last
six years. This is a chain of the thinking that eventually lead
to the {\it final} Standard Model \cite{Hwang417}.

We do know that the world of the minimal Standard Model, the visible
ordinary-matter world, seems to be extremely simple. One originally
starts out with the electron, a point-like spin-1/2 particle, and ends
up with other point-like Dirac particles (such as quarks, other
leptons, and so on), but with interactions through gauge fields
modulated by the Higgs fields. This is what we experimentally know, and
it is a little strange that it seems to be "complete" and that nothing
else seems to exist (until the dark-matter world "calls for" the
"extended" Standard Model). Note that all particles in the minimal
Standard Model are "point-like" since under the best resolution of
$10^{-20}\, cm$ they don't seem to have a size. After Dirac's equation
we have in fact searched for the point-like particles, now for over eighty
years. It seems that Dirac equations explain all the relativistic point-like
particles and their interactions. So, why don't we formulate a working rule to
describe this fact? Let's call it as "the Dirac Similarity Principle".

The other striking experimental fact of the minimal Standard Model is that,
for the past forty years, we have been looking for the Higgs particle(s) -
the spin-0 scalar particle(s), but any signature for the Standard Model
is so far barely seen (July 2012). I try to formulate this fact as
"the minimum Higgs hypothesis". Here we don't mean "the non-existence
of Higgs" but rather we mean that if they exist then they should be minimal.
In our opinion, mass generation in our "basic" theory is so important that
this Higgs question has to be answered one way or the other.

In other words, it would be too broad to identify which extended Standard
Model could be the choice. With the "minimum Higgs hypothesis", the Higgs
sector is essentially fixed once the group $SU_c(3) \times SU_L(2) \times
U(1) \times G$ with $G$ the extension is fixed. On the other hand, the
"Dirac Similarity Principle" helps to fix the particle contents. With these
working rules \cite{Hwang3} that are in essence used for several decades,
the search for the correct extended Standard Model could be sharpened and
thus much easier.

In the minimal Standard Model that has been experimentally verified and that
describes the ordinary-matter world (to a large extent), it could be understood
as a world consisting of a set of point-like Dirac particles interacting
through gauge fields modulated (minimally) by the Higgs fields. The only
unknowns are neutrinos, which, in view of the massive nature, may also be
point-llke 4-component Dirac particles. Thus, the minimal Standard Model is
basically a Dirac world interacting among themselves through gauge fields
modulated by the Higgs fields. In extending the minimal Standard Model, we
try to keep the "principle" of point-like Dirac particles intact - thinking
of the eighty-year experience some sort of sacred. On the other hand, the
forty-year search for Higgs (scalar fields) amounts to the "minimal Higgs
hypothesis". These two working hypotheses simplify a lot of things. In
fact, the situation from something un-manageable without the working rules
becomes manageable, if the two rules are adopted. Of course, we have to
worry about what if one of the two working rules doesn't hold, but it
doesn't hurt to begin in this particular way.

That is, we follow another paper \cite{Hwang3} and introduce "the Dirac
similarity principle" that every "point-like" particle of spin-1/2 could be
observed in our space-time if it is "connected" with the electron, the
original spin-1/2 particle. For some reason, this clearly has
something to do with how relativity and the space-time structure gets married
with spin-1/2 particles. This is interesting since there are other ways to
express spin-1/2 particles, but so far they are not seen
perhaps because they are not connected with the electron. In other words, the
partition between geometry (in numbers such as $4\times 4$ $\sigma/2$ in the
angular momentum) and space-time (such as $\vec r \times \vec p$) is similar
to the electron. We adopt "Dirac similarity principle" as the working
"principle" as we extend the Standard Model to describe the
dark-matter particles as well.

These are "point-like" Dirac particles of which the size we believe is
less than $10^{-20}\,cm$ (assuming that they are point-like up to a few $TeV$
in energy), the current best resolution of the length. Mathematically, the
"point-like" Dirac particles are described by "quantized Dirac fields" - maybe
via a renormalizable lagrangian. The "quantized Dirac fields", which we may
axiomatize for its meaning, in fact does not contain anything characterizing
the size (so far at least). The word "renormalizability" might contain
statements describing the change of the sizes. In other words, "point-like"
is something difficult to describe.

We know that there are three generations of quarks and leptons
but don't know why. This is clearly a symmetry that we have already
"seen", but not in details. On the other hand, neutrino oscillations
are firmly established these days. These oscillations are transitions
from one flavor to another - from one generation to another. This is a
clear evidence for the lepton-flavor-violating interaction
\cite{Family1}. The existence of the cross product in the triplet
$(\nu_\tau,\, \nu_\mu,\, \nu_e)$ (bi-linear, off-diagonal to guarantee
the conversions in oscillations) calls for the existence of the
(complex scalar) dark-matter Higgs triplet, under the family group
$SU_f(3)$. We assume that the scheme is renormalizable, in the naive
power-counting sense. The existence of this extra term is unique.

Coming back to think about it, neutrino oscillations are now
experimentally verified; quantum mechanically, it has to be proceeded
through some lepton-flavor-violating interaction, as above. The
coupling of the Dirac fields to the Higgs field is unique.

The family gauge $SU_f(3)$ theory has to be a massive gauge theory
since those hidden loop diagrams would show up at low energies. This
leads to the family gauge $SU_f(3)$ theory recently proposed
\cite{Family}. Besides the straightforward "derivation" for the
existence of the $SU_f(3)$ gauge theory, we suspect that most
symmetries may be realized in the form of gauge theories. On other hand,
the missing right-handed sector is always a troublesome for the symmetry
people (like me). So, the next option is to consider the idea originated
by Pati and Salam \cite{Salam} that the left-right symmetry might be
restored at some even higher energy.

However, when we construct the Standard Model such as \cite{Hwang417},
we realize the importance of choosing the so-called "basic units" -
one basic unit will come with one kinetic-energy term, and only
one; the idea of Pati and Salam \cite{Salam} would violate this
one-to-one correspondence.

Unfortunately, their idea is in violation of the "gauge principle",
in the sense that the right-handed and left-handed components of
a Dirac particle enter some "basic units", i.e., the multiplets,
{\it once and only once}, and that since the right-handed component
is already in the $SU_L(2)$ singlet, it has to {\it repeat} its
appearance as the $SU_R(2)$ doublet. The gauge principle requires
that there is only one kinetic-energy term which, under chirality,
is split into one right-handed component and one left-handed
component. The right-handed fermion which is a $SU_L(2)$ singlet, if
also regarded as the $SU_R(2)$ doublet, cannot match with a single
kinetic-energy term.  This is why we think that Pati and Salam
\cite{Salam} could not fly.

If the extended Standard Model would be based on the group $SU_c(3)
\times SU_L(2) \times U(1) \times U(1)$, the extra $Z^{\prime 0}$
extension \cite{Hwang}, the extra or "remote" Higgs doublet would be
the only choice. After one degree of freedom gets eaten by the extra
$Z^{\prime 0}$, there remains three degrees of freedom, a little too
much. Since we adopt the "minimum Higgs hypothesis", we'd look for other
options.

In fact, we have a strong case in support of the $SU_f(3)$
family gauge theory \cite{Family}. Neutrino oscillations call for a
lepton-flavor-violating interaction, which must be in the form
${\bar \Psi}_L(3,2) \times \Phi(3,2) \cdot \Psi_R(3,1) + h.c.$, an
off-diagonal (cross-generation) interaction. We need a massive
$SU_f(3)$ family gauge theory - a kind of spontaneous-broken
gauge symmetry. The scenario
from a pair of complex scalar Higgs fields is such that the eight
gauge bosons and the four left-over Higgs particles all are
massive. Under the "minimum Higgs hypothesis", the structure
of the underlying Higgs mechanism is pretty much determined. Then,
neutrinos acquire their masses, to the leading order, with the aid
of both the Higgs $SU_f(3)$ triplets, in fact, from $\Phi(3,2)$
(the mixed family Higgs) and $\Phi(3,1)$ (the purely family
Higgs) \cite{Hwang417, Origin}. Of course, the loop diagrams
involving the gauge bosons also contribute to neutrino masses,
to higher orders.

Thus, we see that the three generations are already there (even
though we have not seen the feeble interactions so far, except
those through the VEV's of the family Higgs). Judging from the
energies which we could reach at LHC, we could set the
scale at a few $TeV's$.

To push forward the "final" minimal extended Standard Model,
we should have a comprehensive successful phenomenology. The
physics of the neutrino sector gets severely modified but
would be very difficult to measure, due to the feeble nature
of the interactions. For example, the lepton-flavor-violating
decays such as $\mu \to e + \gamma$ would be there but they
are beyond reach \cite{Family1}. $\nu_\mu + N \to \tau + N^*$
would become possible on top of the ordinary reaction
$\nu_\mu + N \to \mu + N^*$. And some other reactions in
violation of the $\tau-\mu-e$ universality might be seen
eventually. The importance of these $SU_f(3)$-related
dimensionless neutrino couplings is that the neutrino
is only species in the ordinary matter that acts also
as dark matter - but almost all these related decays
will proceed undetected, owing to the involved
dark-matter particles including neutrinos.
If the $SU_f(3)$ gauge sector is there, its communication
with us (the ordinary matter) is only through the neutrinos.
Things, like the breakdown of the $\tau-\mu-e$ universality,
albeit it could be very small, is absolutely crucial.

Qualitatively, dark-matter particles refers to those particles which
do not participate the ordinary strong and electromagnetic interactions.
It is a natural but stringent definition of "the dark matter".
Under this definition, all the particle species in the $SU_f(3)$
sector, as they do not participate {\it directly} the ordinary
strong and electromagnetic interactions, can be classified as
"dark matter". That is, the extensions of the minimal Standard
Model may bring in new ordinary-matter particles and
occasionally some dark-matter particles, such as the above
specific extra $Z^{\prime 0}$ model.

The $SU_f(3)$ family gauge theory is so far the only natural
way to get the new species of dark matter, besides the neutrinos.
The additional other gauge group is rather limited - the model
of Pati and Salam \cite{Salam} serves as a good illustrative
counter-example.

{\it The gauge principle together with the well-known
$SU_c(3) \times SU_L(2) \times U(1)$ multiplets, plus the
constraint of two chiral components for each particle, gives
gives us the important requirement. Under the gauge-principle
requirement that the two chiral components of each particle
can appear {\bf once and only once} to ensure one complete
kinetic-energy term, the additional group has to be
compatible with $SU_c(3)\times SU_L(2) \times U(1)$ -
it should be either $SU_f(3)$, or additional $U(1)$, or
additional $SU_L(2)$, etc. The additional group
must be $SU(3)$, or $SU_L(2)$, or $U(1)$, or some product
of them. $SU(3)$ for the family symmetry is a good
possibility, for the three generations of fermions.
The Standard Model of \cite{Hwang417} is unique, since
the gauge-group structure is already closed and
consistent among themselves. It is closed and complete,
indeed.}

In other words, the gauge group {\it and the existing
multiplets} already determines everything. To proceed
a little further, we may argue that there is no
more model-building, except the trivial ones. Thus,
except those $SU_f(3)$-related dark-matter particles or
similar (even more feeble ones), there is no more
dark-matter particles. All $SU_f(3)$ family gauge
bosons and the associated family Higgs particles, which
are dark-matter particles (in addition to neutrinos),
decay rather rapidly, i.e., they are short-lived
for the problem of galactic formation and
evolution.

\bigskip

\section{The Standard Model is a Dimensionless Theory.}

The Standard Model of particle physics \cite{Hwang417}
is crucial for the problem of galactic formation and
evolution. If we examine the Standard Model in detail
\cite{Origin, Fields}, we realize that, apart from
the "ignition" term (for the SSB), all couplings, in
the "background" (the force fields and the Higgs), the
lepton world, and the quark world, are dimensionless -
i.e., the Standard Model is originated from a complete
dimensionless theory. Furthermore, that it is a
dimensionless theory means that the origin
of these couplings come from the quantum 4-dimensional
Minkowski space-time {\it as a whole}, rather than from
its local field contents.

Conversely, we could start from the (quantum) 4-dimensional
Minkowski space-time with the force-fields gauge-group structure
$SU_c(3) \times SU_L(2) \times U(1) \times SU_f(3)$ built-in
from the outset - then derive the Standard Model, thus determine
all dimensionless couplings. It sounds like that everything is
the consequence of the quantum 4-dimensional Minkowski
space-time - at least we feel that way. As a result, the
problem of galactic formation and evolution should be
synthesized from the Standard Model
\cite{Hwang417, Origin, Fields}.

\medskip

Let's introduce, in some detail, the quark world and then the
lepton world.

In the Standard Model \cite{Hwang417}, we live in the quantum
4-dimensional Minkowski space-time with the force-field
gauge-group structure $SU_c(3) \times SU_L(2) \times U(1)
\times SU_f(3)$ built-in from the outset. This is what we call
"the background".

Then, what is the quark world? The quark world is a world of matter
form, thus of the type of Dirac equations. It claims a rather small
length scale, of about $10^{-13}cm$. The strong-interaction
nature of $SU_c(3)$ explains such small size. The color $SU_c(3)$
gauge fields are already classified as part of "the background" -
the quarks of three colors and of six flavors are building blocks
of matter for the quark world. The quark world knows the gauge
group $SU_c(3) \times SU_L(2) \times U(1)$, but not $SU_f(3)$ --
the so-called (123) symmetry.

The quark world knows color $SU_c(3)$ well - the strong interaction
that acts in the range of fermi's (i.e., $10^{-13}\, cm$). Everything
larger than a few fermi would eventually cut off the influence of
the strong interaction, unless some special arrangements are given
(by the God).

The lepton world is very similar except that the scale is much
bigger, at the atomic scale, or $10^{-8}\, cm$. But it does not
know the color $SU_c(3)$, except indirectly.

"Our world" is the combination of the background, the quark
world, and the lepton world - so, it is quite complicated
but in fact all of them are (quantum) point-like particles.
Amazingly, they could be represented as a branch of
mathematics - or, relativistic quantum mechanics and
quantum field theory \cite{Book}.

The decomposition of the Standard Model could make our thoughts
much clearer, eventually to adopt a language which is precise
\cite{definition}. Such as: we live in the (quantum) 4-dimensional
Minkowsi space-time with the force-fields gauge-group structure
$SU_c(3) \times SU_L(2) \times U(1) \times SU_f(3)$ built-in
from the outset - as the "background" of our world.
This background supports the quark world. For some reason, it also
supports the lepton world.

In introducing the family concept as a gauge group, we
regard \cite{HwangYan} $((\nu_\tau,\,\tau)_L,\,(\nu_\mu,\,\mu)_L,
\,(\nu_e,\,e)_L)$ $(columns)$ ($\equiv \Psi(3,2)$) as the
$SU_f(3)$ triplet and $SU_L(2)$ doublet.
It is essential to complete the (extended) Standard Model \cite{Hwang417}
by working out the Higgs dynamics in detail \cite{Origin}. It is also
essential to realize the role of neutrino oscillations - it is the change
of a neutrino in one generation (flavor) into that in another generation;
or, we need to have the coupling $i h \bar \Psi_L(3,2)\times
\Psi_R(3,1) \cdot \Phi(3,2)$, exactly the coupling introduced by Hwang
and Yan \cite{HwangYan}. Then, it is clear \cite{Hwang417} that the
mixed family Higgs $\Phi(3,2)$ must be there. The remaining purely
family Higgs $\Phi(3,1)$ helps to complete the picture, so that
the eight gauge bosons are massive in the $SU_f(3)$ family gauge
theory \cite{Family}.

Thus, we see $SU_f(3)$ in the lepton world but it seems that
the quark world does not see $SU_f(3)$ at all. Maybe in the
quark world the $SU_f(3)$ forces are much too feeble than the
$SU_c(3)$ forces.

Another point might be critical. In the quark world, we cannot get
a hand on the mass of an individual quark, because we cannot see
an isolated quark. So, the issue might be how, for us, to reach
the meaning of "mass" for a composite system, such as a
three-quark system, the quark-antiquark system, etc.

Remember that the story is pretty much fixed if the so-called
"gauge-invariant derivative", i.e. $D_\mu$ in the kinetic-energy
term $-\bar \Psi \gamma_\mu D_\mu \Psi$, is given for a given
basic unit \cite{Book}. It seems that this aspect is as fundamental
as the Einstein relation, $E^2={\vec p}^2+m^2$.

Thus, we have, for the up-type right-handed quarks $u_R$, $c_R$,
and $t_R$,
\begin{equation}
D_\mu = \partial_\mu - i g_c {\lambda^a\over 2} G_\mu^a -
i {2\over 3} g'B_\mu,
\end{equation}
and, for the rotated down-type right-handed quarks $d'_R$, $s'_R$,
and $b'_R$,
\begin{equation}
D_\mu = \partial_\mu - i g_c {\lambda^a\over 2} G_\mu^a -
i (-{1\over 3}) g' B_\mu.
\end{equation}

On the other hand, we have, for the $SU_L(2)$ quark doublets such as
$(u_L,\,d'_L)$,
\begin{equation}
D_\mu = \partial_\mu - i g_c {\lambda^a\over 2} G_\mu^a - i g
{\vec \tau\over 2}\cdot \vec A_\mu - i {1\over 6} g'B_\mu.
\end{equation}

That is, we are using $d'_R$, $d'_L$, etc., consistently. In the
quark world, the down-type quarks are always rotated - that means
that the so-called GIM are always there.

The mass term from the old Standard-Model way is given by
\begin{eqnarray}
L_m&=& -G_1 \{ {\bar d}'_R \Phi^\dagger(1,2) Q_{1,L} + h.c. \}
       -G'_1 \{ {\bar s}'_R {\tilde \Phi}^\dagger(1,2) Q_{1,L} +h.c. \}\nonumber\\
&&   - G_2 \{ {\bar s}'_R \Phi^\dagger(1,2) Q_{2,L} + h.c.\}
     - G'_2 \{ {\bar c}_R {\tilde \Phi}^\dagger(1,2) Q_{2,L} + h.c. \}\nonumber\\
&&   - G_3 \{ {\bar b}'_R \Phi^\dagger(1,2) Q_{3,L} +h.c.\}
     - G'_3 \{ {\bar t}_R {\tilde \Phi}^\dagger(1,2) Q_{3,L} + h.c. \}.
\end{eqnarray}
Note that the six couplings $G_{1,2,3}$ and $G'_{1,2,3}$
in principle can be adjusted. The unitary mixings, the so-called
GIM mechanism \cite{GIM}, for the down-type quarks helps to
forbid the weak neutral current, at the expense of introducing
peculiar cross-mass terms. How to detect these "peculiar"
interactions through the SM Higgs studies would be
something of importance and urgency.

Assuming that the $u$ and $d$ be massless ($G_1\sim G'_1 \sim 0$),
we have to assume four parameters (couplings) to generate four
masses. It seems possible to take $m(t)/m(b) = m(c)/m(s)$, so
then there is some (hidden) symmetry.

In general, the rotation among the down-type quarks means that
there are off-diagonal masses such as $m_{ds}$, etc. In fact,
this is something which we cannot avoid, because $G_1$, $G_2$,
and $G_3$ are three different mass parameters.

In any event, we so far cannot say too much for the "individual"
quark masses for the quark world. It seems that the GIM mechanism,
while cutting of the neutral weak currents, introduces the
off-diagonal mass terms, in a more hidden part of the story.

\medskip

We could follow the following logic in reaching the
global picture. Neutrino oscillations tell us that there
exists an interaction, such as $i {\bar \Psi}_L \times
\Psi_R \cdot \Phi$, where the scalar field and the
left-handed and right-handed fermions are $SU_f(3)$
triplets. We know that the left-handed fermion has to
be the doublet under $SU_L(2)$ - that pushes the
scalar field $\Phi$ also an doublet under $SU_L(2)$.
Thus, we have an interaction of the form,
$i {\bar \Psi}_L(3,2) \times \Psi_R(3,1) \cdot
\Phi(3,2) + h.c.$.

Since we put all six objects as a representation
in the group, we agree that all these objects are
point-like Dirac particles - {\it not} the mixture
of Dirac particles and Majorana particles.
Moreover, particles of the second or third
generation must be of the same characteristics
as the particles of the first generation. All these
are the group theory in mathematics - we physicists
sometime forget the mathematics ABC.

Of course, we are too far in proving experimentally
that these neutrinos are also point-like Dirac
particles. The regularities, or the symmetries, sort
of give us the confidence in all this regarding the
Standard Model.

In the lepton world, we introduce the family triplet,
$(\nu_\tau^R,\,\nu_\mu^R,\,,\nu_e^R)$ (column), under $SU_f(3)$.
Since the minimal Standard Model does not see the right-handed
neutrinos, it would be a natural way to make an extension of the
minimal Standard Model. Or, we have, for $(\nu_\tau^R,\,
\nu_\mu^R,\,\nu_e^R)$,
\begin{equation}
D_\mu = \partial_\mu - i \kappa {\bar\lambda^a\over 2} F_\mu^a.
\end{equation}
and, for the left-handed $SU_f(3)$-triplet and $SU_L(2)$-doublet
$((\nu_\tau^L,\,\tau^L),\, (\nu_\mu^L,\,\mu^L),\, (\nu_e^L,\,e^L))$
(all columns),
\begin{equation}
D_\mu = \partial_\mu - i \kappa {\bar\lambda^a\over 2} F_\mu^a - i g
{\vec \tau\over 2} \cdot \vec A_\mu + i {1\over 2} g' B_\mu.
\end{equation}
The right-handed charged leptons form the triplet $\Psi_R^C(3,1)$ under
$SU_f(3)$, since it were singlets their common factor $\bar\Psi_L(\bar 3,2)
\Psi_R(1,1)\Phi(3,2)$ for the mass terms would involve the cross terms such as
$\mu\to e$.

The neutrino mass term assumes the {\it unique} form:
\begin{equation}
i {h\over 2} {\bar\Psi}_L(3,2) \times \Psi_R(3,1) \cdot \Phi(3,2)
+ h.c.,
\end{equation}
Here the Higgs field $\Phi(3,2)$ is the mixed family Higgs, because
it carries some nontrivial $SU_L(2)$ charge. In fact, the charged
part of $\Phi(3,2)$ does not experience the spontaneous
symmetry breaking (SSB), as worked out explicitly in \cite{Origin}.

We wish to note, again, that, for charged leptons, the Standard-Model
choice is $\Psi^\dagger(\bar 3,2) \Psi_R^C(3,1) \Phi(1,2) +c.c.$, which
gives three leptons an equal mass. But, in view of that if
$(\phi_1,\phi_2)$ is an $SU(2)$ doublet then $(\phi_2^\dagger,
-\phi_1^\dagger)$ is another doublet, we could form
${\tilde\Phi}^\dagger(3,2)$ from the doublet-triplet $\Phi(3,2)$.

\begin{equation}
i {h^C\over 2} {\bar\Psi}_L(3,2) \times \Psi_R^C(3,1) \cdot
{\tilde \Phi}^\dagger(3,2) + h.c.,
\end{equation}
which gives rise to the imaginary off-diagonal (hermitian) elements
in the $3\times 3$ mass matrix, so removing the equal masses of the
charged leptons.

In the quark world, the GIM mechanism via the down-type quark
mixings, such as Eq. (5) in the previous section, seems to
explain the quark masses - with the six couplings $G_{1,2,3}$
and $G'_{1,2,3}$. In contrast, the lepton world has only
two couplings $h$ and $h^C$, controlling the neutrinos
and the charged leptons - $h$ would be very small in
light of the tiny neutrino masses, and $h^C$ about the $\tau$
or $\mu$ mass. For both the quark world and the lepton
world, these are dimensionless couplings; the same for
the force-fields $SU_c(3) \times SU_L(2) \times U(1) \times
SU_f(3)$ built-in from the outset gauge-group structure.

Because everything is dimensionless, it has to do with
the 4-dimensional Minkowski space-time; it has nothing
to do with the individual fields. Hence, this story is
absolutely beautiful - the act of the Einstein relation
and of its Dirac's linearization in the quantum
4-dimensional Minkowski space-time.

\medskip

Suppose that, before the spontaneous symmetry breaking (SSB), the Standard Model
does not contain any parameter that is pertaining to "mass", but, after the SSB,
all particles in the Standard Model acquire the mass terms as it should - we
call it "the origin of mass". In what follows, we wish to show this indeed
the case - explaining the origin of mass \cite{Origin}. In this way, we try
to tie "the origin of mass" to the effects of the SSB, or the generalized
Higgs mechanism.

We start at the Higgs sector of the Standard Model \cite{Hwang417}. To begin
with, we have the Standard-Model Higgs $\Phi(1,2)$, the purely family Higgs
$\Phi(3,1)$, and the mixed family Higgs $\Phi(3,2)$, with the first label
for $SU_f(3)$ and the second for $SU_L(2)$. We need another triplet $\Phi(3,1)$
since all eight family gauge bosons are massive \cite{Family}.

In the U-gauge, we choose to have
\begin{equation}
\Phi(1,2)= (0,{1\over \sqrt 2} (v+\eta)),\,\, \Phi^0(3,2) = {1\over
\sqrt 2} (u_1+\eta'_1, u_2+ \eta'_2, u_3+\eta'_3 ),\,\,
\Phi(3,1) = {1\over \sqrt 2}(w+\eta',0,0),
\end{equation}
all in columns. The five components of the complex triplet $\Phi(3,1)$ get
absorbed by the $SU_f(3)$ family gauge bosons and the neutral part of
$\Phi(3,2)$ has three real parts left - together making all eight family
gauge bosons massive.

In this way, we must write \cite{Origin}

\begin{eqnarray}
V_{Higgs} =& \mu^2_2 \Phi^\dagger(3,1) \Phi(3,1) + \lambda
(\Phi^\dagger(1,2) \Phi(1,2)+ cos\theta_P\Phi^\dagger(3,2)\Phi(3,2))^2\nonumber\\
    &  + \lambda(-4 cos\theta_P)
(\Phi^\dagger(\bar 3,2)\Phi(1,2))(\Phi^\dagger(1,2)\Phi(3,2))
  \nonumber\\
  &+\lambda
(\Phi^\dagger(3,1) \Phi(3,1)+ sin\theta_P \Phi^\dagger(3,2)\Phi(3,2))^2
  \nonumber\\ &  + \lambda(-4 sin\theta_P)
(\Phi^\dagger(\bar 3,2)\Phi(3,1))(\Phi^\dagger(3,1)\Phi(3,2)).
\end{eqnarray}
These are two prefect squares minus the other extremes, to guarantee
the positive definiteness, when the minus $\mu^2_2$ was left out.
($\theta_P$ may be referred to as "Pauchy's angle".)

One important point: The "ignition" point is on the purely
family Higgs $\Phi(3,1)$, rather than on the SM-Higgs
$\Phi(1,2)$. It yields the "prediction", $m_{SM\,\,Higgs}
=v/2$ with $v$ the SM Higgs vacuum expectation value (VEV).

Another important point: All the terms have the universal
dimensionless coupling $\lambda$. We have $\lambda={1\over
8}$, a self-repulsive interaction.

The last critical point: Apart from the "ignition" term,
the Higgs sector is dimensionless. This fact, together
with what are in the lepton world and in the quark world,
implies that the Standard Model \cite{Hwang417} is a
dimensionless theory - all the couplings are determined
by the (quantum) 4-dimensional Minkowski space-time
{\it as a whole}, rather than its local contents,
i.e., its fields, etc.

\bigskip

\section{Clustering during the young age of our Universe}

The problem of galactic formation and evolution has a lot of
meaning if we could start from the Standard Model of particle
physics \cite{Hwang417}. This is the main focus of the
present paper.

Essentially, we try to "prove" the Standard Model from
the various angles. We declare that we live in the quantum
4-dimensional Minkowski space-time with a specific
force-fields gauge-group structure built-in in the very
beginning - the "background" of our world. Luckily, we can
"see" the lepton world and we also can "see" the quark
world. We should have enough of confidence in using this
Standard Model \cite{Hwang417, Origin, Fields} to study
the problem of galactic formation and evolution.

Suppose that the spiral of the Milky Way is caused by the dark-matter aggregate of
four or five times the mass of the Milky Way, and similarly for other spiral galaxies.
This aspect will serve as a "basic fact" for our analysis of the problem of galactic
formation and evolution.

Based on the Standard Model \cite{Hwang417}, we have to identify
the dark-matter aggregate associated with the Milky Way as
a halo of neutrinos and antineutrinos. The heavy dark-matter
particles are short-lived family gauge bosons (familons) and
family Higgs, and no others - the lifetime of a few $keV$ or
less ($\sim 10^{-20}\,sec^{-1}$). That is why, assuming the
Standard Model \cite{Hwang417}, the identification of the
dark-matter aggregate as a huge neutrino halo is quite natural.

Let's look at our ordinary-matter world. Those quarks can aggregate in no time, to
hadrons, including nuclei, and the electrons serve to neutralize the charges also
in no time. Then atoms, molecules, complex molecules, and so on. These serve as
the seeds for the clusters, and then stars, and then galaxies, maybe in a time span
of $1\, Gyr$. The aggregation caused by strong and electromagnetic forces is fast
enough to give rise to galaxies in a time span of $1\, Gyr$. On the other hand, the
weak interactions proceed fairly slowly in this time span and they could not
contribute in the time span of $1\, Gyr$. The said aggregation in the
ordinary-matter world gave rise to the "seeded" clustering, so much faster
than if there would be no "seeded" clustering (presumably, so many order of
magnitude away).

The clustering, if for the moment we are allowed to look at it from Newton's
gravitation law, requires some seed, i.e., some attraction center, to begin
with. Then, the time needed for a gravitational system to shrink to a half,
or others, can be computed relatively easily. The important point is
that, under Newton's gravitational law, an object has to have a mass
to be movable. For our model galaxies, the huge neutrino encircling
halo has the visible ordinary-matter seed, such as the spiral galaxy,
in the construction, since there is no dark-matter seed for the
system of purely neutrinos/antineutrinos. Under the same token,
there is no photon sphere - i.e., no clustering for photons.
There exists one importance evidence - the $3^\circ\,K$ cosmic
microwave background (CMB) is rather uniform.

For the clustering of CB$\nu$'s, the decoupling happened a little
earlier than the $3^\circ\,K$ CMB and we may assume that the
patterns, including its time dependence, of the clustering of
neutrinos follow the ordinary-matter chunks, since
Newton's gravitational law would give a universal acceleration
independent of the object mass. The ordinary-matter chunks,
those formed from protons, etc., eventually became stars, and,
later, galaxies. Unlike CMB, the clustering of CB neutrinos makes
them non-uniform and, as said earlier, the neutrinos halos
would distribute following the visible stars and the visible
galaxies. This is one place where Newton's gravitational
law might still hold reasonably.

The birth of the first galaxy probably happened at
the time $1\,Gyr$ after the Big Bang. We agree that
we live in the quantum 4-dimensional Minkowski space-time
with the force-fields gauge-group structure built-in from
the very beginning - as the "background" of "our world".
Thus, we see the lepton world and also we see the quark
world. The aggregation on matters leads to the various
"seeded" clusterings. The clustering for neutrinos,
with any ordinary-matter seeds, follows the pace of
the visible (ordinary-matter) chunks, as we think that
Newton's gravitational law still applies.

{\it Formation of model galaxies is the destiny, with
the visible parts being the spiral, the elliptical, the
interacting, or the others, and the much larger
invisible neutrino halos. Because now we know that
at least one species of neutrinos/antineutrinos
has the mass $\sim 0.058\,eV$ and it would be
that neutrinos/antineutrinos are the final matter-matter
decay products, the model galaxies, such as the Milky
Way, the Satellites, and much bigger invisible
neutrino clouds, have to precede their final corpses.}

\bigskip

\section{Summary: Partially dark-matter galaxies}

{\it Our world is comprehensible, after all, in physics
as well as in mathematics. At the end of the 20th Century,
we firmly establish that we in fact live in the quantum
4-dimensional Minkowski space-time with the force-fields
gauge-group structure built-in from the very beginning,
with the CMB and CB$\nu$ everywhere. In the 21st Century,
we should move forward, taking for granted the strange
world which we live and solving the problems as such.}

Granting the validity of the Standard Model \cite{Hwang417,
Origin, Fields}, the problem of galactic formation and
evolution is solved in that the galaxies would be in the
form of the model galaxy plus the variants. {\it The model
galaxy is composed of two parts, the much larger enclosing
invisible neutrinos/antineutrinos part, and the visible
part consisting of ordinary matter such as the Milky Way
and its Satellites.}

We believe that, for the clusterings, there is no
"seed" from the dark-matter world - equivalently,
there is no long-lived dark-matter heavy particle
that could serve as the seed of the clustering. The
Standard Model is so complete and the world coming
out of it is rather simple.

\bigskip

\section*{Acknowledgments}
This research was supported in part by National Science Council project (NSC
99-2112-M-002-009-MY3).


\bigskip

\begin{thebibliography}{99}

\bibitem{Hwang417} W-Y. Pauchy Hwang, arXiv:1304.4705v2 [hep-ph] 25 August
2015.

\bibitem{Origin} W-Y. Pauchy Hwang, The Universe, {\bf 2-2}, 47 (2014).

\bibitem{Fields} W-Y. Pauchy Hwang, The Universe, {\bf 3-1}, 3 (2015).

\bibitem{Hwang2} W-Y. Pauchy Hwang, arXiv:1012.1082v6 [hep-ph] 13 Jan 2016.

\bibitem{PDG} Particle Data Group, "Review of Particle Physics",
J. Phys. G: Nucl. Part. Phys. {\bf 37}, 1 (2010); and its biennial
publications.

\bibitem{definition} W-Y. Pauchy Hwang, arXiv:1409.6077v1 [hep-ph] 22 Sep 2014;
A precise definition of the Standard Model.

\bibitem{Hwang3} W-Y. P. Hwang, arXiv:11070156v1 [hep-ph] 1 Jul 2011); Plenary
talk given at the 10th International Conference on Low Energy Antiproton Physics
(Vancouver, Canada, April 27 - May 1, 2011).

\bibitem{Family1} W-Y. Pauchy Hwang, arXiv:1207.6443v1 [hep-ph] 27 July 2012.

\bibitem{Family} W-Y. Pauchy Hwang, Nucl. Phys. {\bf A844}, 40c (2010);
W-Y. Pauchy Hwang, International J. Mod. Phys.
{\bf A24}, 3366 (2009); the idea first appeared in
hep-ph, arXiv: 0808.2091; talk presented at 2008 CosPA Symposium
(Pohang, Korea, October 2008), Intern. J. Mod. Phys. Conf. Series {\bf 1}, 5
(2011); plenary talk at the 3rd International Meeting on Frontiers of Physics,
12-16 January 2009, Kuala Lumpur, Malaysia, published in American Institute of
Physics 978-0-7354-0687-2/09, pp. 25-30 (2009).

\bibitem{Salam} J.C. Pati and A. Salam, Phys. Rev. {\bf D10}, 275 (1974);
R.N. Mohapatra and J.C. Pati, Phys. Rev. {\bf D11}, 566 (1975); {\bf D11},
2559 (1975).

\bibitem{Hwang} W-Y. P. Hwang, arXiv:1009.1954v2 (hep-ph, 27 May 2011);
W-Y. P. Hwang, Phys. Rev. {\bf D36}, 261 (1987); see the second paper for
more earlier references.

\bibitem{Book} Ta-You Wu and W-Y. Pauchy Hwang, "Relativistic Quantum
Mechanics and Quantum Fields" (World Scientific, 1991), currently
being updated into the second edition.

\bibitem{HwangYan} W-Y. Pauchy Hwang and Tung-Mow Yan, The Universe,
Vol. 1, No. 1, 5 (2013); arXiv:1212.4944v1 [hep-ph] 20 Dec 2012.

\bibitem{GIM} S.L. Glashow, J. Iliopoulos, and L. Maiani, Phys. Rev.
{\bf D2}, 1285 (1970).


\end{thebibliography}
\end{document}